\documentclass[11pt,a4paper]{article}

\usepackage[utf8]{inputenc}   
\usepackage[T1]{fontenc}          
\usepackage{graphicx}                   
\usepackage{float}                      % Mieux positionner les figures
\usepackage{geometry}                   % Pour les marges du document
\usepackage{hyperref}                   % Pour les hyperliens
\usepackage{setspace}                   % Pour l'interligne
\usepackage{color}                      % Pour pouvoir utiliser les couleurs
\usepackage[table]{xcolor}              % Couleurs dans les tableaux
\usepackage{fancyhdr}                   % Pour entêtes/pieds de pages
\usepackage{fourier-orns}               % Pour utiliser les symboles /!\ etc

\hypersetup{                            % Configuration des hyperliens
        colorlinks=true,                % Colorise les liens
        breaklinks=true,                % Retour ligne liens trop longs
        urlcolor=black,                 % Couleur des hyperliens
        linkcolor=black,                % Couleur des liens internes
        pdftitle={},                    % Titre du PDF
        pdfauthor={O. Bailleux},        % Auteur du PDF
        pdfproducer={O. Bailleux},      % Producteur du PDF
        pdfsubject={PPC},               % Sujet du PDF
        pdftoolbar=true,                % Barre d'outils Acrobat
        pdfmenubar=true,                % Menu Acrobat
        pdfstartview={FitH},            % ouvrir PDF largeur de la page
}

\sffamily                               % Police de caractères
\title{\vspace{-1cm}Reified unit resolution and the failed literal rule}

\normalsize\author{Olivier Bailleux, Université de Bourgogne}

\begin{document}

\sloppy % Toujours respecter la marge de droite 

%Entete de pages
%\renewcommand{\chaptermark}[1]{\markboth{\chaptername~\thechapter{} : #1}{}}
%\renewcommand{\sectionmark}[1]{\markright{\thesection{} #1}}
%\renewcommand{\headrulewidth}{0.4pt}
%\renewcommand{\headsep}{24pt}
%\setlength\headheight{34pt}
%\headwidth = \textwidth
%\lhead{\small\includegraphics[width=3mm]{images/book_open.png} \leftmark}
%\lhead{O. Bailleux - 2010}
\rfoot{\small page \thepage / \pageref{pageend}}

\maketitle 

%\parskip=1pt
%\begin{spacing}{1}
%\tableofcontents
%\end{spacing}
%\parskip=10pt

%\clearpage

% Pour avoir un interligne de 1,1
\begin{spacing}{1.05}

\begin{abstract}
Unit resolution can simplify a \textsc{cnf} formula or detect an inconsistency by repeatedly assign the variables occurring in unit clauses.
Given any \textsc{cnf} formula $\sigma$, we show that there exists a satisfiable \textsc{cnf} formula $\psi$ with size polynomially related to the size of $\sigma$ such that applying unit resolution to $\psi$ simulates all the effects of applying it to $\sigma$. The formula $\psi$ is said to be the \emph{reified counterpart} of $\sigma$.
This approach can be used to prove that the failed literal rule, which is an inference rule used by some \textsc{sat} solvers, can be entirely simulated by unit resolution. More generally, it sheds new light on the expressive power of unit resolution.
\end{abstract}

\section{Introduction}

\subsection{Unit resolution}

A unit clause is a logical clause with only one literal, like $(a)$ or $(\overline b)$. Unit resolution (also called unit propagation) consists to repeatedly fix the variables occurring into unit clauses in such a way to satisfy these clauses. For example, if there is a clause $(\overline b)$ in the formula, then the variable $b$ is set to \texttt{false}, and the formula is simplified by removing all the clauses containing $\overline b$ as well as all the occurrences of $b$ in the other clauses. 

Sometimes, unit propagation produces the empty clause, meaning that the formula is not satisfiable. Because unit resolution in not a complete proof system, all unsatisfiable formulae cannot be solved in this way. In \textsc{sat} solvers, unit propagation is used to fixe some variables in order to reduce the number of branches in the search tree.

\subsection{The failed literal rule}

This is an inference rule allowing \textsc{sat} solvers to fix some variables which cannot be fixed by using only unit propagation. As an example, let us consider the following \textsc{cnf} formula:
\begin{equation}
\label{eq}
\sigma = (a \vee b) \wedge (\overline{b} \vee c) \wedge (\overline{b} \vee \overline{c})
\end{equation}

Because there is no unit clause, applying unit resolution to this formula does not fix any variable. Applying the failed literal rule to the literal $\overline{a}$ consists in \emph{trying} to fix this variable to \texttt{false} and then to apply unit propagation. Because the empty clause is produced, $\sigma \wedge \overline{a}$ is not satisfiable. Then the variable $a$ \emph{must} be set to \texttt{true}. The failed literal rule can be also applied to the literal $b$, with the result that the variable $b$ must be set to \texttt{false}.

\subsection{Contribution}

We will show that applying unit propagation to a \textsc{cnf} formula $(\overline{l} \vee \overline{w}) \wedge \mathtt{reif}( \sigma \wedge w, l)$ has the same effect as applying the failed literal rule to a formula $\sigma$ with the literal $w$. $\sigma' = \mathtt{reif}( \sigma \wedge w, l)$ is said to be the \emph{reified counterpart} of $\sigma \wedge (w)$ in the sense that applying unit propagation to $\sigma'$ cannot produce the empty clause, but fixes $l$ to \texttt{true} if and only if applying unit propagation to $\sigma \wedge (w)$ would produce the empty clause.

Although the size of the reified counterpart $\mathtt{reif}(\psi, l)$ of a formula $\psi$ is polynomially related to the size of $\psi$, the interest of the concept is rather theoretical. It sheds new light on the expressive power of unit resolution.

\section{Reified unit resolution}

The unit propagation process can be decomposed into several steps, where each step $i$ fixes the variables which occur in unit clauses after the step $i-1$ (if applicable) is completed. Because each step fixes at least one variable, and because the empty clause is produced when the same variable is fixed both to \texttt{true} and \texttt{false}, the number of steps cannot exceed $n+1$, where $n$ is the number of variables in the formula. Let $\sigma$ be a \textsc{cnf} formula with $n$ variables, and $\psi$ its reified counterpart. The formula $\psi$ can be decomposed in $n+1$ sub-formulae $\psi_1, \ldots, \psi_{n+1}$, where each $\psi_i$ simulates the effect of the step $i$ of unit propagation on $\sigma$. For each variable $v$ of $\sigma$, there are $2(n+1)$ variables, namely $v_1^+, v_1^-, \ldots, v_{n+1}^+, v_{n+1}^-$, in $\psi$. The formula $\psi$ is designed so that if $v$ is fixed to \texttt{true} (\texttt{false}, respectively) after $i$ propagation steps on $\sigma$, then $v_i^+$ ($v_i^-$, respectively) is fixed to \texttt{true} after $i$ propagation steps on $\psi$. As a manner of speaking, the assignations $v = \mathtt{true}$ and $v = \mathtt{false}$ are decoupled in $v_i^+ = \mathtt{true}$ and $v_i^- = \mathtt{true}$ in $\psi$, and no variable of $\psi$ can be set to \texttt{false} by unit propagation. 

Let us present the construction of $\psi$ form the formula
\begin{equation}
\sigma = (\overline{a}) \wedge (a \vee b) \wedge (\overline{b} \vee c) \wedge (\overline{b} \vee \overline{c})
\end{equation}

The sub-formula $\psi_1$ must allow unit propagation to fix $a_1^-$ to \texttt{true} because at the first step of unit propagation on $\sigma$, the variable $a$ is fixed to \texttt{false}. Then

\begin{equation}
\psi_1 = (a_1^-)
\end{equation}

The sub-formula $\psi_2$ must allow unit propagation to fixe $a_2^-$ to \texttt{true} because $a$ remains to \texttt{true} at the second step of unit propagation on $\sigma$. This can be obtained thanks to the clause $(\overline{a_1^-} \vee a_2^-)$, which will be called a \emph{propagation clause}.\ It must also allow unit propagation on $\psi$ to simulate the effect of unit propagation on $\sigma$ regarding the clause $(a \vee b)$, given that $a$ is set to \texttt{false}. This can be obtained thanks to the clause $(\overline{a_1^-} \vee b_2^+)$, which will be called a \emph{deduction clause}. 

Because the goal is to build the formula $\psi$ without knowing in advance which variables of $\sigma$ will be fixed by each unit resolution step, all the possible propagation and deduction clauses are added to each sub-formula $\psi_i, i>1$.

\begin{equation}
\begin{array}{ccl}
\psi_i & = & \overbrace{(\overline{a_i^-} \vee a_{i+1}^-) \wedge (\overline{a_i^+} \vee a_{i+1}^+) \wedge (\overline{b_i^-} \vee b_{i+1}^-) \wedge (\overline{b_i^+} \vee b_{i+1}^+) \wedge (\overline{c_i^-} \vee c_{i+1}^-) \wedge (\overline{c_i^+} \vee c_{i+1}^+)}^{\mathrm{propagation\ clauses}}\\
 & \wedge & \underbrace{(\overline{a_i^-} \vee b_{i+1}^+) \wedge (\overline{b_i^-} \vee a_{i+1}^+) \wedge (\overline{b_i^+} \vee c_{i+1}^+) \wedge (\overline{c_i^-} \vee b_{i+1}^-) \wedge (\overline{b_i^+} \vee c_{i+1}^-) \wedge (\overline{c_i^+} \vee b_{i+1}^-)}_{\mathrm{deduction\ clauses}}\\
\end{array}
\end{equation}

For example, the third propagation clause $(\overline{b_i^-} \vee b_{i+1}^-)$ says "if $b_i^- = \mathtt{true}$ at the step $i$ of unit propagation on $\psi$, meaning that $b = \mathtt{false}$ at the step $i$ of unit propagation on $\sigma$ then $b_{i+1}^-$ must be set to \texttt{true}  at the step $i+1$ of unit propagation on $\psi$, meaning that $b = \mathtt{false}$ at the step $i+1$ of unit propagation on $\sigma$".

As another example, the third deduction clause $(\overline{b_i^+} \vee c_{i+1}^+)$ says "According to the clause $(\overline{b} \vee c)$ of $\sigma$, if $b_i^+ = \mathtt{true}$ at the step $i$ of unit propagation on $\psi$, meaning that $b = \mathtt{true}$ at the step $i$ of unit propagation on $\sigma$, then $c_{i+1}^+$ must be set to \texttt{true}  at the step $i+1$ of unit propagation on $\psi$, meaning that $c = \mathtt{true}$ at the step $i+1$ of unit propagation on $\sigma$".

The production of the empty clause by unit propagation on $\sigma$ (if applicable) can be reified by adding a new variable $s$ and the following clauses to $\psi$

\begin{equation}
(\overline{a_4^+} \vee \overline{a_4^-} \vee s) \wedge (\overline{b_4^+} \vee \overline{b_4^-} \vee s) \wedge (\overline{b_4^+} \vee \overline{b_4^-} \vee s)
\end{equation}

Clearly, unit propagation on $\psi$ will fix $s$ to \texttt{true} if and only if unit propagation on $\sigma$ produces the empty clause, i.e. implicitly fixes the same variable both to \texttt{true} and \texttt{false}.

As it stands, the formula $\psi$ is of little interest because it can only allow to simulate one "scénario" of unit propagation on $\sigma$. It is much more useful to simulate the effects of unit propagation when some variables of $\sigma$ have been previously fixed (for example by other inference rules or by branching rules in the context of the running of a \textsc{sat} solver).

To this end, some of (or all) the variables of $\sigma$ can be injected into $\psi$ with the following clauses:

\begin{equation}
(\overline{a} \vee a_1^+) \wedge (a \vee a_1^-) \wedge (\overline{b} \vee b_1^+) \wedge (b \vee b_1^-) \wedge (\overline{c} \vee c_1^+) \wedge (c \vee c_1^-)
\end{equation}

Thanks to these additional clauses, unit propagation on $\psi$ can simulate the effect of unit propagation on $\sigma$ under any given partial truth assignment of the variables of $\sigma$.

If $\sigma$ includes $n$ variables and $m$ clauses with at most $k$ literals per clause, then each sub-formula $\psi_i$ contains $2n$ binary propagation clauses and at most $km$ $k\mathtt{-ary}$ deduction clauses. It follows that $\psi$ contains $O(n^2+nkm)$ clauses.

\section{Concluding remarks}

We shown that for any formula $\sigma$, there exists a \emph{satisfiable} formula $\psi$ such that unit propagation on $\psi$ can simulate the behavior of unit propagation on $\sigma$, even when the empty clause is produced. What this tells about the expressive power of unit propagation ? Unit propagation can be seen as a way to compute functions mapping partial truth assignments to $\{ \mathtt{yes}, \mathtt{no} \}$ with two different approaches. In the first approach, the result \texttt{yes} corresponds to the assignment of a particular variable. In the second one, it corresponds to the production of the empty clause. The results presented in this report show that the \emph{same functions} can be computed using these two approaches, and that the required numbers of clauses are polynomially related.

\label{pageend}

%\begin{center}
%\includegraphics[scale=0.5]{graphe.eps}
%\end{center}

% Pour finir l'interligne de 1,1
\end{spacing}

\end{document}